# Time dynamics of photothermal vs opto-acoustic response in mid-IR nanoscale biospectroscopy

Peter D. Tovee[a)], Claire Tinker-Mill[a)], Kevin Kjoller[e)], David Allsop[b)], Peter Weightman[c)], Mark Surman[d)], Michele R. F. Siggel-King[c, d)], Andy Wolski[c)], and Oleg V. Kolosov[a)*]

a) Physics Department, Lancaster University, Lancaster, LA1 4YB, UK,
b) Division of Biomedical and Life Sciences, Lancaster University, Lancaster, LA1 4YB, UK,
c) Department of Physics, University of Liverpool, Oxford street, Liverpool, L69 3BX, UK,
d) Science and Technology Facility Council, Sci-Tech Daresbury, Keckwick Lane, Daresbury, Warrington, WA4 4AD, UK,
e) Anasys Instruments, 325 Chapala Street, Santa Barbara, CA 93101, US.

*Corresponding author; o.kolosov@lancaster.ac.uk, +44 (0)1524 593619, www.nano-science.com

**Abstract**

Infrared (IR) spectroscopy, a well-established tool for chemical analysis of diverse materials, has significant potential in biomedical applications. While the spatial resolution of traditional IR spectroscopy is limited by the wavelength of the IR light to the few micrometres, it has been shown that nanoscale chemical analysis can be obtained by detecting IR induced local heating photo-thermal response via Scanning Thermal Microscopy (SThM) or local thermomechanical expansion using Atomic Force Microscopy (AFM). This paper explores the potential of a pulsed ps pulse duration high power free electron laser (FEL) light source for AFM-IR and SThM-IR spectroscopy employing standard AFM and SThM probes. The SThM-IR response was found to have a detrimental strong background signal due to the direct heating of the probe, whereas the AFM-IR thermomechanical response allowed to eliminate such a problem for both top-down and bottom-up illuminations with the FEL IR source. The SThM-IR characteristic response time was approximately half that of AFM-IR, in line with finite element analysis simulations. Finally, the advantages and drawbacks of AFM-IR wavelength sensitive spectroscopic response using a ps-duration FEL vs a high repetition quantum cascade laser IR source in studies of nanoscale dimension amyloid peptide fibres were explored both experimentally and via finite elements analysis.

## 1. Introduction

IR spectroscopy in the 2.5-20 µm (mid-IR) wavelength range is of great interest for chemical identification of organic compounds, biological samples and polymer materials due to the characteristic spectroscopic absorption lines of molecular vibrations falling in this range. While Fourier transform IR microscopic spectroscopy (FT-IR) has been used for many decades for characterization of such objects, its spatial resolution is intrinsically confined to few micrometres by the light wavelength limiting the dimension of the spot IR radiation can be focused into. This in turn severely impedes the ability of FT-IR to investigate objects with sub-µm features, such as polymeric nano-composites and living cells, and to foster the development of nano-sensors - all of which being important components of a rapidly expanding nanotechnology field. One approach to bring the lateral resolution of IR microscopy to the nanoscale is to use scanning probe microscopy (SPM) [1-6]. While SPM measurements are often affected by tip-sample geometry and require relatively flat samples, they were nevertheless successfully used for probing diverse sample properties [7, 8], including local IR absorption [9-12] using the AFM-IR approach of measuring the photothermal (PT) induced expansion of the sample [13, 14]. Alternatively, scanning thermal microscopy (SThM) can be used for detection of IR absorption, directly related to the local heating of the sample [15, 16]. This has the additional advantage of simultaneously detecting the photothermal expansion of the sample. In the past Wollaston wire probes have been used to detect the sample heating in FT-IR based photothermal microspectroscopy (PTMS) [17, 18], however these Wollaston wire probes have a larger tip radius than the more modern SThM probes with the lateral resolution on the μm length scale. Initial AFM-IR experiments used a FEL as the IR laser source, with quantum cascade lasers (QCL) [19, 20] and optical parametric oscillators (OPO) being the more recent choice for IR light sources [21, 22]. Both QCL and OPO offer the ability to tune or sweep through a range of wavelengths similar to the FEL but without the need for a large accelerator facility. QCLs can provide sufficient average power necessary for thin, less absorbing samples.

AFM-IR exploits the photothermal effect [13], where a tuneable IR laser heats a thin sample and an AFM tip detects the thermally induced local expansion of the sample at the tip-surface contact, usually only a few nm in size. The incident laser wavelength is tuned to within one of the sample's IR absorption bands with the partially absorbed light causing vibrational excitation of the sample molecules that rapidly decays into heat. This heat causes sample expansion which is detected as a deflection of the AFM cantilever in contact with the sample surface. As a result, if the IR wavelength is swept, the AFM signal reflects the specific absorption lines of the sample. The IR light can be pulsed (as with FEL, OPO or some QCL sources) or modulated at sub-MHz frequency with QCL sources. With the pulsed source, the AFM cantilever receives an initial kick that can be detected via Fourier Transform (FT) of the cantilever 'ring down' [14]. For the modulated source, the frequency of the modulation is usually tuned to the contact resonance frequency of the cantilever, strongly enhancing the AFM-IR signal [23]. The approach has also been shown to work in a liquid environment opening the way for nanoscale IR spectroscopy of living cells [24]. However while the dimensions of the tip-sample contact are on the few nm scale, AFM-IR lateral resolution also depends on the local heat transfer and elastic deformation phenomena effectively reducing resolution to a few tens of nanometres [10, 11, 24]. AFM-IR also can operate with either top-down or bottom-up laser illumination. In top-down mode, the laser is shone directly onto the sample surface making it easier to select the imaging area. This has the disadvantage that the laser will also hit the AFM tip causing photothermal expansion of the tip and cantilever itself that can mask the useful photothermal signal from the sample. This effect can be reduced by Au coating of the tip that would reflect the IR light and this has the additional benefit of enhancing the useful signal due to the tip-related field enhancement effect [5, 19]. Bottom-up AFM-IR requires an IR transparent prism with high refractive index (usually ZnSe) to illuminate the sample via the evanescent field wave [12]. While this approach requires fairly thin samples (of not more than few μm) that have to be deposited on the prism, and also makes positioning of the laser beam and the probe more complicated, it has the advantage that the probe itself is not directly heated by the incident IR beam, resulting in a more clean spectroscopic signal that depends on the absorption of the sample alone.

Spectrally sensitive information in AFM-IR can be acquired either at a single point by sweeping the IR wavelength, or as a sequence of AFM raster scan images obtained at several selected wavelengths. Fixed point spectra allow for chemical analysis similar to FT-IR with AFM-type resolution, whereas fixed wavelength AFM images allow for certain features to be highlighted by tuning the laser wavelength to a known absorption of the molecule or structure. In this work we mostly used fixed wavelength raster scans comparing photothermally induced AFM-IR mechanical response with thermal measurements using SThM probes. The objects used were polymer based PMMA/graphite mixure as well as Aβ-42 amyloid fibres [25, 26], with the latter being the key target testing the feasibility of AFM-IR pulsed FEL approach for biomedical applications. This paper makes direct comparisons between AFM-IR deflection and SThM-IR deflection using SThM thermal signals for polymer and Aβ amyloid fibre samples using FEL-IR light source. The experimental data of nanomechanical and nano-thermal response are supported by the theoretical finite element analysis (FEA) simulations. Amyloid fibres data obtained with the FEL are also compared with the commercial Anasys “NanoIR2” system that uses a pulsed QCL.

## 2. Materials and Methods

***2.1. Optical and electronic setup.*** Two different lasers were used during this study; the ALICE (Accelerators and Laser In Combined Experiments) FEL IR laser source (Daresbury Laboratory, UK), and a visible laser (Stradus, Vortran Technology) operating at 637 nm with peak power up to 140 mW, with the latter mostly used to determine time-dependent optothermal and cantilever dynamics phenomena in AFM-IR. The ALICE FEL laser was tuneable over approximately 5 to 9 µm range producing a “macropulse” at a repetition rate of 10 Hz and duration of 85 µs, with average power up to 10 mW. Each ALICE macropulse consists of approximately 1400 micropulses of approximately 1 ps duration and a peak power of 0.85 MW, firing at a repetition rate of 16.25 MHz. The 637 nm Stradus laser allowed digital modulation with frequencies up to 200 MHz, allowing us to emulate repetition rate and a total energy of a macro and micro-pulse structure of ALICE.

$CaF_2$ flat disks and ZnSe prisms (Crystran Ltd, Poole, UK) were used as IR transparent substrates. The ZnSe prisms allowed for top down as well as bottom up laser illumination from internal reflection. The parallel surfaces of $CaF_2$ discs permitted top down only illumination in our setup.

The experimental setup for AFM-IR and SThM-IR is shown in Fig. 1 (see also supporting information Fig. S1 for the optics setup). A Multi-Mode SPM with Nanoscope IIIa controller (Bruker, USA) was used for all the measurements, with Contact-G probes (Budget Sensor, Bulgaria) used in AFM-IR mode, and SThM $Si_3N_4$ probes (Kelvin Nanotechnologies) used for both AFM-IR and SThM-IR measurements [27-29]. For the AFM-IR mode an in-house signal access module breakout box was used to take the deflection signal from the AFM that was amplified by the band-pass filter (SR650, Stanford Research Systems, USA) to amplify the signal, with the high pass filter set to reduce a background thermal signal noise. The boxcar averager (Delta Developments, UK) triggered by the FEL macropulse signal was then used to record the signal and to feed it into the AFM controller which then simultaneously recorded the FEL intensity and the wavelength. The SThM-IR measurements employed the SThM $Si_3N_4$ probes forming part of the Maxwell-Wheatstone electrical bridge. The bridge was excited by the combined DC offset (providing probe Joule self-heating during thermal imaging) and 91 kHz AC signal (used to measure the probe resistance), with resistive and capacitive elements of the bridge allowed nulling bridge output at low AC amplitude and absence of DC offset when self-heating of the probe can be neglected [28, 30, 31]. Balancing of the bridge circuit was done in ambient environment and with no FEL or AFM deflection laser applied with the tip away from the sample surface. The output of the bridge was connected to the 90 kHz band-pass filter (SRS650, Stanford Research Systems, USA) that also acted as a differential amplifier for the bridge output (see figure 1). In the SThM-IR mode of scanning only a DC offset resulting in probe current of *ca* 1 mA was applied to the SThM bridge using a function generator (33500B waveform generator, Agilent, USA) [28, 29].

IR light from the FEL was directed to the sample via Au-coated mirrors and a $CaF_2$ lens providing an illuminated area of approximately 50x20 μm on the sample surface. The FEL's auxiliary HeNe red laser collinear with the IR beam was used as a positioning guide to optimise the optical alignment. Finally, the AFM-IR signal was maximised by adjusting the mirror tilt immediately before scanning. In order to change between the top-down and bottom-up illumination in this setup, a simple parallel displacement of the beam was required with minor shift of lens position along the optical axis. The intensity of the FEL beam ($I_0$) was monitored using a single-element pyro detector (THZ21-BL-BNC, Gentec, USA) on a portion of the beam split off using a $CaF_2$ beam splitter, and subsequently used for ratiometric AFM-IR measurements. Additionally, the analogue signal proportional to the IR wavelength value was provided independently from the FEL system wavelength monitor software and hardware.

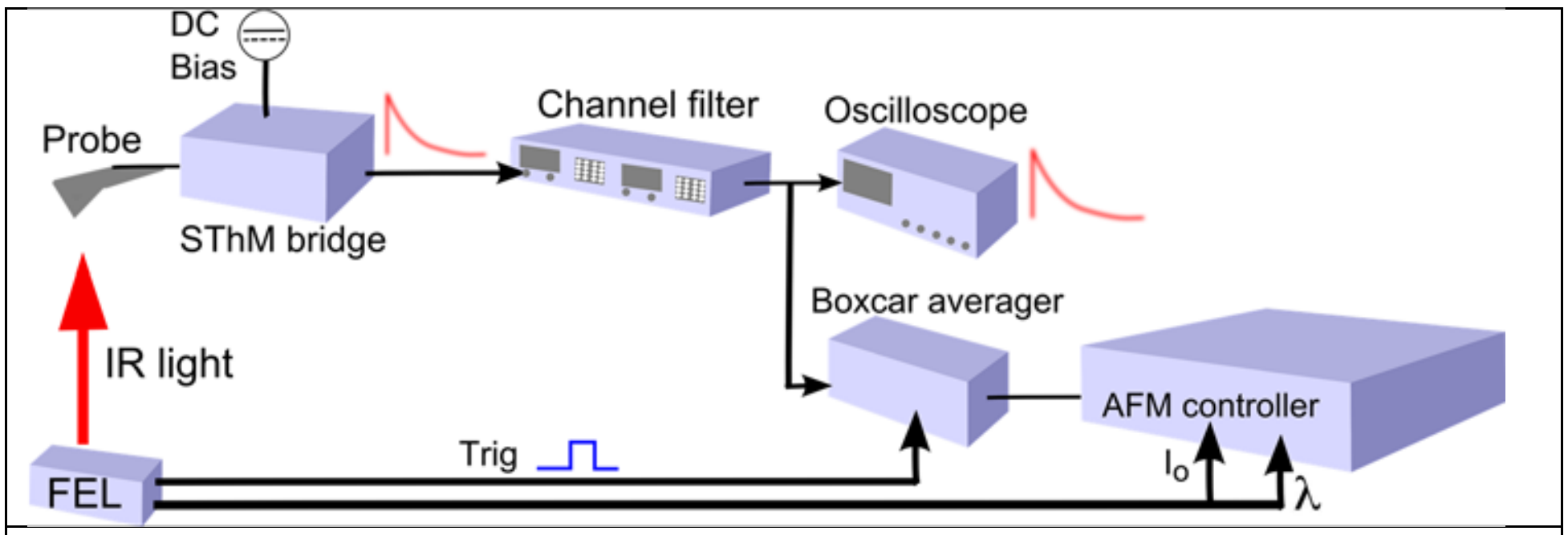

**Fig 1.** AFM-IR/SThM-IR experimental setup for ALICE IR laser, where $I_0$ is proportional to the IR intensity from the pyro detector and $\lambda$ is the wavelength signal from the FEL system.

***2.2. Sample preparation.*** Samples were deposited on ZnSe prisms and $CaF_2$ discs. ZnSe and $CaF_2$ substrates were cleaned by sonication in acetone and IPA for 15 minutes in each solvent before being washed with deionized (DI) water and blown dry with pure nitrogen immediately before deposition of the samples. Graphite particles dispersed in a PMMA matrix were used as test samples since both graphite and PMMA absorb in the IR wavelength range used. The test sample was made by dispersing graphite soot in PMMA photoresist solution (A4, MicroChem, USA) by sonication for 15 minutes followed by drop casting the solution on a ZnSe prism, followed by curing for 3 minutes on a hot plate at 100° C. The ZnSe prisms were secured to a metal AFM disc at the side keeping the central part of the back surface of the prism free which was essential for the bottom-up illumination of the sample.

The amyloid fibre samples were prepared from the recombinant human Aβ1:42 samples (Ultra-pure, HFIP, A-1163-2, >97 % purity, rPeptide, Georgia, USA). The peptide was deseeded using a protocol adapted from Manzoni *et al.*[32] and separated from the smaller aggregates via centrifugation (see detailed protocol in Supporting Information, section 1.1). Once the amyloid in solution was prepared it was pipetted onto $CaF_2$ discs and ZnSe prisms and then dried. The amyloid fibre samples were stored in a desiccator with silica gel to keep them reasonably dehydrated as hydration may affect the IR absorption signature of the sample.

***2.3. FEA simulation.*** 3D finite element simulations were performed using the software package COMSOL Multiphysics©. The transient sample heating by the IR laser pulse and the resulting thermomechanical response was simulated. The expansion was coming from the thermal heating of the sample (PMMA or amyloid fibre) with the much stiffer substrate modelled as a fixed constraint. A time dependant solver was used to solve dynamic response over a time domain of 100 times that of the heating pulse. Since the thermomechanical properties of amyloid fibres are not known, the characteristics of a typical polypeptide polymer were used as an approximation. In the FEA simulations the PMMA sample was modelled as a 200x400x100 µm block thermally anchored on the bottom and sides (see figure 3). The heated area of the PMMA samples was set as 250x25x5 µm reflecting the typical focusing of the tilted FEL beam and the IR absorption depth in the material. The amyloid fibres were simulated as a 50 nm diameter and 500 nm length rod on a 10x10x5 µm ZnSe substrate. Due to the long fibre length and therefore large contact area, the thermal boundary resistance between the fibre and substrate was not a significant factor and not considered in the models. The pulse duration was approximately 85 µs and the peak laser power in the macropulse on the surface was 0.1 W for both the PMMA and Aβ fibres samples. The resulting heat power density for PMMA was 3.2 $\times10^{12}$ W $m^{-3}$ and for Aβ fibres 4 $\times10^{15}$ W $m^{-3}$.

## 3. Results and discussion

### *3.1. Main characteristics and time dependence of the AFM-IR and SThM-IR responses with FEL IR excitation.*

When a section of a material is heated, the heat spreads out until temperature equilibrium is reached. The characteristic time for the process is defined by the heat diffusion time $t_c = {z^2}/{D}$, where $z^2$ is the distance heat travels and $D$ the material diffusivity [33]. The characteristic size of the amyloid fibres is determined by their diameter that was typically 50 nm for the fibre bunches, whereas for the PMMA, the characteristic size was defined by the laser spot size at about 50 µm. This allows to estimate characteristic times as 28 ns and 25 ms for the amyloid fibre and PMMA, respectively. The FEL macropulse duration was 85 µs, which is much longer than the characteristic time for the fibre. Therefore heating of the amyloid fibre can be regarded as being well within the quasistatic regime.

The maximum temperature $T$ of the sample can be estimated from the following equation after [33]

$$T = \frac{P_D w}{K\sqrt{2\pi}}$$

where $K$ is the sample thermal conductivity, $w$ the sample profile width (assumed to be about 50 nm for amyloid fibres) and $P_D$ the power density of the incident light (dependant on laser spot size), which gives a temperature increase of approximately 50 K. Given the thermal expansion $\Delta L = \Delta T L \alpha$, where $L$ is the original size and $\alpha$ the thermal expansion coefficient, the expansion was estimated to be on the order of 0.2 nm. While small, it is well within the sensitivity range of the AFM for the tip deflection.

For PMMA the characteristic time was much longer than the heat pulse therefore this falls within the dynamic heating regime. The maximum surface temperature will then also depend on the pulse duration and can be estimated using [33]

$$T(\rho, z, t) = \frac{\emptyset_0 w^2}{C_V(\pi D t)^{1/2}(8Dt + w^2)} \exp\left(\frac{2\rho^2}{8Dt + w^2} - \frac{z^2}{4Dt}\right)$$

where $D$ the thermal diffusivity, $t$ time since pulse, $w$ the profile width, $C_V$ the sample heat capacity, $\emptyset_0$ the thermal flux and $\rho$ the position within the laser spot, with $\rho$ for the centre of the laser beam being 0. By integration over the 85 µs laser pulse and for the assumption of an infinite PMMA sample, the expansion was estimated to be 38 nm. This was 3 orders of magnitude larger than the amyloid fibres, as would be expected given the relative sizes of the samples and heated areas.

In the experiment, we first tested the AFM dynamic response to optothermal heating using the 637 nm visible laser light which allowed for greater variability of time domain excitation to explore the fundamentals of the AFM and SThM-IR techniques. We observed both the dynamic cantilever deflection and SThM thermal response via oscilloscope traces as shown in Fig. 2.

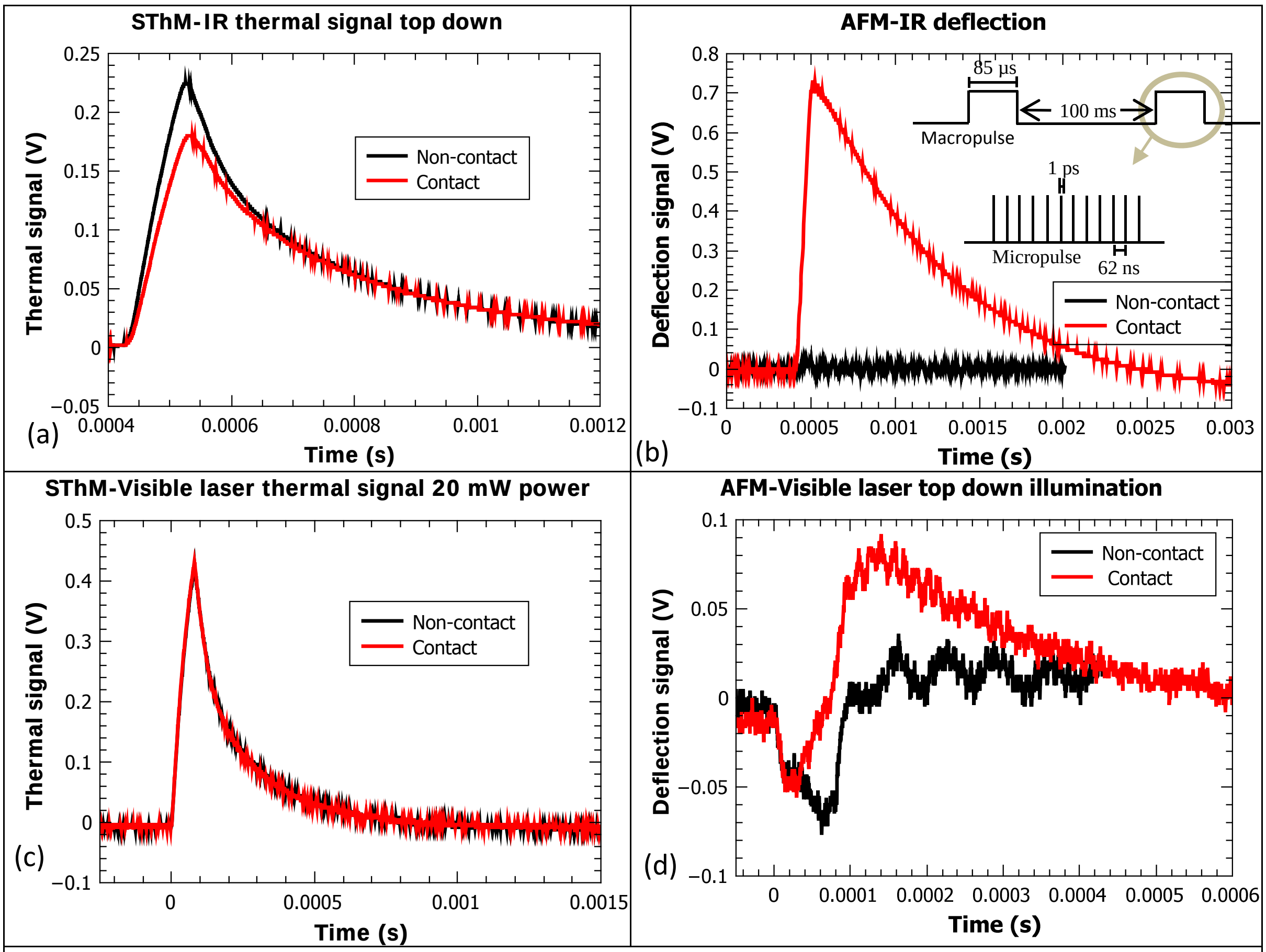

**Fig 2.** Time domain response of; SThM-IR (thermal) signal cantilever response (a) with the (b) AFM-IR cantilever deflection signal for the FEL IR laser illumination, and the SThM signal (c) for the visible laser excitation as well as (d) the AFM deflection response deflection response from the optomechanical oscillations driven by the visible laser. The inset (b) shows the macro and micro-pulse structures.

Figure 2b and Fig. 2d shows that for the AFM-IR mode there are still small oscillations when the AFM cantilever is out-of-contact with the sample at a frequency of approximately 15 kHz, the resonance frequency of a free cantilever. It can be observed that the thermal response of SThM-IR to the FEL laser pulse was lower when the tip was in contact, compared to the non-contact position, suggesting that the tip is heated to higher temperatures due to both larger heat dissipation in the sample, and, possibly, lower light absorption in the sample. As mentioned above, that would create a large background signal that does not carry information about the sample. As could be expected, the deflection signal of the SThM probe in contact and out-of-contact with the sample are similar to signals obtained with the AFM-IR probe, with the non-contact SThM deflection slightly higher. This is due to the complex bi-metallic structure of the SThM cantilever causing bending when heated. More details of the response are presented in supplementary Fig. S2. In particular, Fig. S2a shows that for the tip in-contact with the sample in absence of FEL beam the noise background is much smaller than the AFM-IR deflection signal. Both top-down and bottom-up laser illumination in this case gave the same deflection signal oscillation pattern.

It should be noted that the signal depends strongly on the focusing and positioning of the laser spot with even small variations in the beam position leading to notable changes in signal intensity. The main origin of this variation is the complex spatial structure of the FEL laser beam, which is also varied with the change of the lasing wavelength. For both SThM and AFM cantilevers the response they received from the sample thermal expansion was sufficiently small to keep the tip in contact with the sample, as can be seen from Fig. 2.

As expected, the responses of the SThM and AFM cantilevers from the visible Stradus laser and the IR FEL were similar, suggesting a possibility to use the easier controlled visible laser to emulate various regimes of AFM-IR and SThM-IR response. In particular, the results confirm that both deflection and thermal response is proportional to the laser power, and changing repetition frequencies within one decade producing the same signal shape. It should be noted, though, that elongation of the heating pulse changed the shape of the ringing pattern, highlighting the significance of the pulse duration over pulse repetition rate at the typical parameters used.

We clearly observed that the characteristic time scale of the SThM-IR thermal signal was much shorter than the timescale for AFM-IR deflection. Using exponential fit, the time constant for the SThM-IR thermal signal was $2.074 \times 10^{-4} \pm 8 \times 10^{-7}$ s and for the AFM-IR deflection signal was $8.35 \times 10^{-4} \pm 2 \times 10^{-6}$ s. This is similar to the SThM probe deflection decay time of $4.3 \times 10^{-4} \pm 2 \times 10^{-5}$ s. The heating decay time was much shorter than the mechanical decay time which is consistent with the results of FEA modelling. The SThM cantilevers are approximately 150 µm long compared to the AFM contact cantilevers which are 450 µm, resulting in a longer mechanical decay time for the AFM cantilevers. The decay time of the sample after the arrival of the laser pulse also depends on the area heated which was less for the smaller amyloid fibres than for the PMMA polymer, and hence the PMMA samples gave a longer decay time.

The results of the FEA simulations (see details in section 2.3.) of laser heating of the sample and the thermomechanical response are presented in Fig. 3. The simulations model both the heat propagation in the sample Fig. 3a,b as well as calculate the force applied to the tip due to sample thermal expansion $F_h$ and the amount of the sample surface expansion, $\Delta Z_h$. Given that the cantilever can be represented as a linear spring with the spring constant $k_c$, these results can be used to numerically calculate the actual observed tip response, $\Delta Z_R$, of the AFM cantilever using the following equation

$$\Delta Z_R = \frac{F_h}{k_c + \frac{F_h}{\Delta Z_h}} \qquad \text{(Eq 1)}$$

Typical values of $k_C$ for AFM cantilevers of 0.2 $Nm^{-1}$, and for SThM probes 0.3 $Nm^{-1}$ were used.

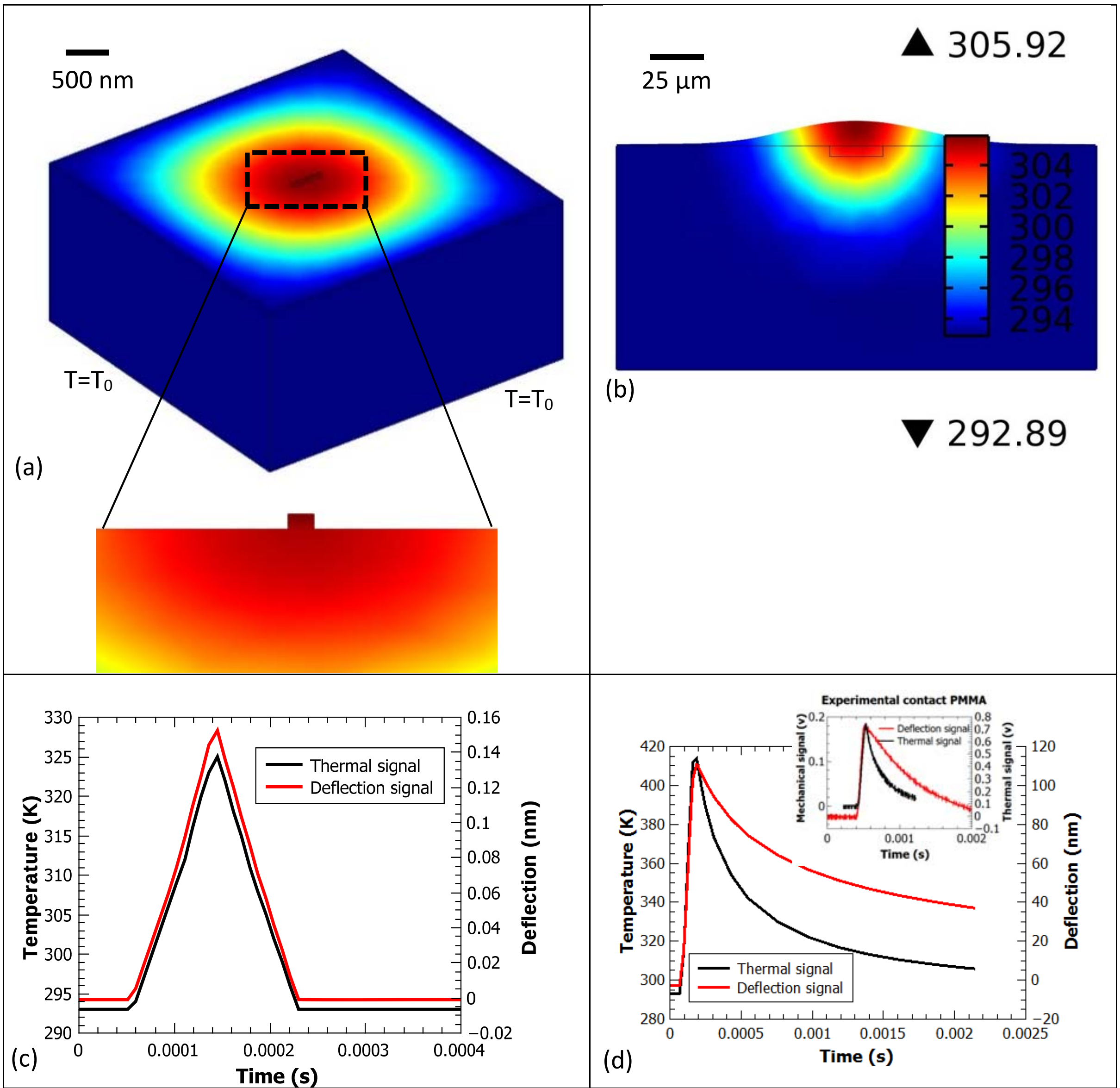

**Fig 3.** Results of the FEA simulations of the temperature distributions in (a) the 50 nm diameter amyloid fibre with zoomed in profile, (b) PMMA polymer showing thermal expansion and graphs of simulated thermal and mechanical signal responses for (d) an amyloid fibre and (d) the PMMA with an inset for comparison with the experimental results.

The modelled PMMA sample signal decay time for the cantilever deflection was longer than the thermal signal decay time, supporting the experimental data. For the amyloid fibre on the non-absorbing substrate, the time dependence of deflection and thermal signals are similar as expected for the much smaller heated volume where the fibre dissipates heat faster into the substrate. That also results in the higher peak temperature for the PMMA sample even though the PMMA and amyloid have thermal conductivities of similar magnitude. As the amyloid fibres and PMMA thermal expansion coefficients are of the same order of magnitude, expansion of the larger heated area of PMMA sample results in an order of magnitude higher displacement compared to the amyloid fibre.

Using exponential decay fitting of the FEA simulation results the decay time of the deflection for the AFM cantilever was 6.3 ± 0.1 x10$^{-4}$ s and for the SThM thermal signal 3.7 ± 0.1 x10$^{-4}$ s, which were commensurate with the the experimental data.

***3.2. Spectroscopic response of the Aβ peptide fibres in AFM-IR mode.*** To explore the spectrally selective response in AFM-IR a standard AFM raster scan was performed on the amyloid fibres on $CaF_2$ substrates. The slow scan was stopped when the fast scan continuing across the fibre, resulting in multiple identical line scans were performed over the resulting fibre section (Fig. 4a). Several AFM-IR line scans were then carried out for fixed wavelengths of 1650 cm$^{-1}$, 1610 cm$^{-1}$ and 1660 cm$^{-1}$ which correspond to absorptions of Amide I, β-sheet and α-helix respectively, Fig. 4b (see also supporting information Fig. S3 for the related raw and averaged data). The results were processed with MATLAB where 128 identical line scans were averaged and normalised by the intensity signal simultaneously gathered with the FEL provided data during this experiment. Fig. 4b shows clearly different AFM-IR signals profiles of Amide I, β-sheet and α-helix across the selected line.

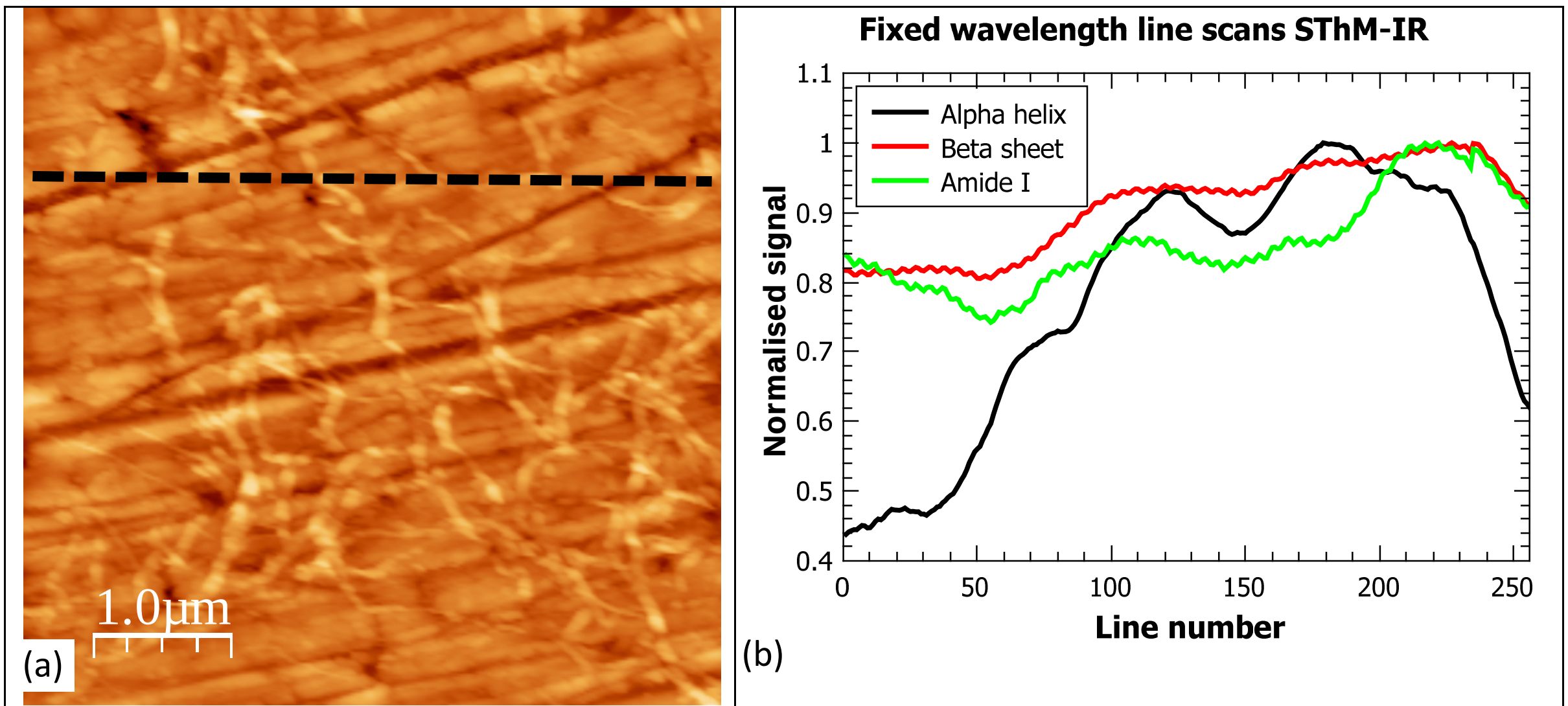

**Fig 4.** Topographical height image (a) of Aβ amyloid fibres on $CaF_2$ with a dotted line showing where the profiles were taken and (b) normalised line profiles for fixed wavelength line scans at; Amide I 1650 cm$^{-1}$, β-sheet 1610 cm$^{-1}$ and α-helix 1660 cm$^{-1}$.

Samples of densely packed Aβ amyloid on Au coated substrate were also studied in the AFM "NanoIR2" commercial system (Anasys Instruments, USA) with results presented in Fig. 5. The NanoIR2 system uses top down illumination with a QCL making it particularly suitable for thin sample with less absorption such as the amyloid fibres.

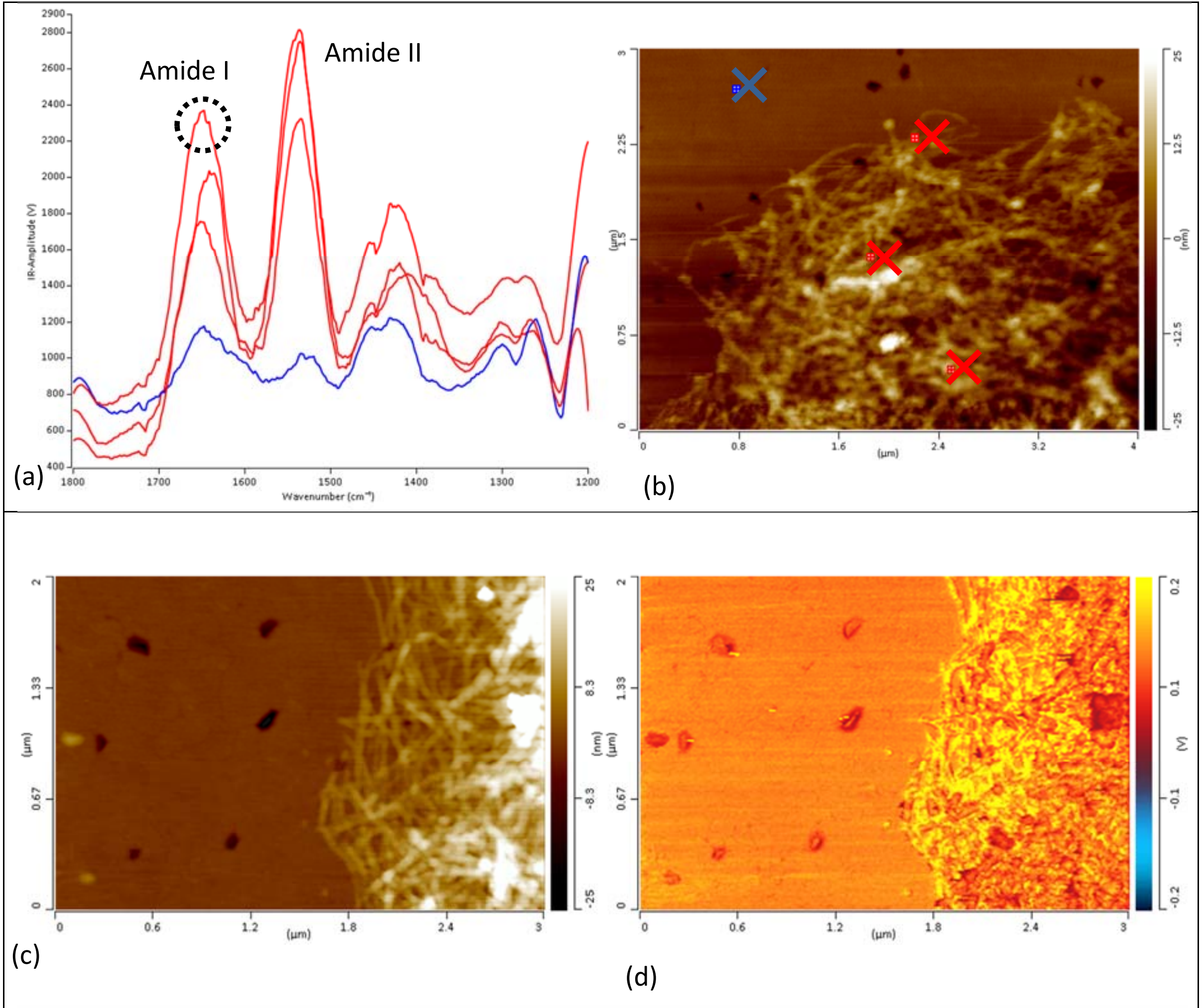

**Fig 5.** NanoIR2 measurement of amyloid fibres on gold substrate showing; (a) spectral absorption data, (b) AFM image of the positions of the spectra, (c) AFM height image for fixed wavelength scan and (d) a full AFM scan done at a fixed wavelength of 1650 $cm^{-1}$ (Amide I). Dotted circle in (a) marks the water absorption effect.

Figure 5a shows strong absorption peaks at approximately 1650 and 1550 $cm^{-1}$ corresponding to Amide I, Alpha helix and Amide II absorption in the amyloid fibre [34-39]. The blue baseline shows the substrate response which shows some contribution at those wavenumbers possibly due to residual surface organic contamination or tip absorption. Nevertheless, that signal had no sharp spectral features and significantly smaller than the amyloid fibre response. Spectra taken on a gold substrate (Fig. S4c) by sweeping the wavelength showed that there was some spectral response even on gold with some dependence on the position of the IR laser spot. Wavelength resolution within the absorption peaks was not good enough to resolve the Amide I and β-sheet wavelengths. It well be possible that the absorbed water molecules interfered with the Amide I line producing a double peak (dotted circle in Fig. 5a). Flushing the system with dry nitrogen would eliminate this feature. One of the biggest problems with IR spectroscopy within the 5-9 µm range is signal from water molecules on the sample surface. Unfortunately this AFM-IR setup did not allow for vacuum or dry $N_2$ measurements which would have removed the effect of water on the spectral response.

The NanoIR2 system was also used to take AFM-IR raster scans at fixed wavelengths; 1650, 1610, 1630, 1696 and 1628 $cm^{-1}$ (Fig. 5d). The results showed some enhancement of the signal caused by the proximity of the AFM tip to the gold substrate due to enhanced scattering. The higher signal was clearly coming from the amyloid fibres compared to the Au coated substrate (see also Fig. S4 in the supporting information for raw spectral data).

## 4. Conclusions

In conclusion, the possibility of using a FEL IR source with complex time-domain pulse structure for thermal SThM-IR and thermomechanical AFM-IR response of polymer and Aβ amyloid fibre samples was investigated. By comparing the dynamic response of optomechanical AFM-IR and optothermal SThM-IR signals it was shown that the SThM-IR characteristic thermal signal decay time was approximately 50 % shorter than the AFM-IR cantilever deflection decay time. These findings are in line with FEA simulations, which showed that the PMMA polymer film had significantly longer characteristic response decay time for both AFM-IR and SThM-IR detection: on the order of 25 µs compared to the amyloid fibre decay time of 28 ns. This was due to much faster heat transport in the small dimension of the amyloid fibres.

Thermomechanical AFM-IR response was shown to have a very small background due to absence of the direct heating of the cantilever as evidenced by comparing in-contact and out-of-contact response for AFM cantilevers. For SThM cantilevers measuring AFM-IR deflection, while possible, resulted in higher thermomechanical out-of-contact response due to complex bi-metallic structure of the cantilever. Therefore the traditional AFM cantilever is best suited for AFM-IR measurements unless simultaneous thermal measurements are required.

Spectroscopically sensitive spatially resolved measurements were performed using line scans over the amyloid fibre samples at fixed wavelengths of 1650 $cm^{-1}$, 1610 $cm^{-1}$ and 1660 $cm^{-1}$ which relate to absorption bands of Amide I, β-sheet and α-helix respectively using the FEL setup. These results were compared with the commercial NanoIR2 system by Anasys Instruments that uses a QCL IR source. The fixed wavelength results showed clear differences for the different wavelengths highlighting either Amide I, β-sheet and α-helix. NanoIR2 spectra also displayed strong peaks for Amide I and II for Aβ fibres compared with the gold substrate. It was also possible to resolve true nanoscale features of the Aβ fibre with the AFM-IR method.

While the pulsed FEL IR source was feasible for AFM-IR measurements, improving the power and spatial stability of the ALICE FEL would further increase the quality of the data. These issues are currently being addressed by ALICE accelerator scientists. Future experiments at ALICE FEL could involve use of the micropulse structure by mixing with AFM ultrasonic excitation at adjacent frequencies and heterodyne detection at the beating frequency. With the micropulse period on the order of 80 ns that would allow further exploration of the thermomechanical response dynamics of amyloid fibres that we estimated in FE simulation to be on the order of 30 ns.

**Acknowledgements**

Authors acknowledge support of EPSRC grant (EP/K023349/1) - Disease diagnosis through spectrochemical imaging (SCANCAN). OVK acknowledges support of EU FUNPROB grant (PIRSES-GA-2010-269169). PDT and OVK would like to thank the STFC ASTeC department for providing access to ALICE and their technical expertise. PDT and OVK would also like to thank Paul Bassan, Tim Craig and James Ingham for the setting up and commissioning of the wavelength monitor and IR end-station.

**Time dynamics of photothermal vs opto-acoustic response in mid-IR nanoscale biospectroscopy**

Peter D. Tovee[a], Claire Tinker-Mill[a], Kevin Kjoller[e], David Allsop[b], Peter Weightman[c], Mark Surman[d], Michele Siggel-King[c], Andy Wolski[c], Oleg V. Kolosov[a]

a) Physics Department, Lancaster University, Lancaster, LA1 4YB, UK,
b) Division of Biomedical and Life Sciences, Lancaster University, Lancaster, LA1 4YB, UK,
c) Department of Physics, University of Liverpool, Oxford street, Liverpool, L69 3BX, UK,
d) Science and Technology Facility Council, Sci-Tech Daresbury, Keckwick Lane, Daresbury, Warrington, WA4 4AD, UK,
e) Anasys Instruments, 325 Chapala Street, Santa Barbara, CA 93101, US.

Corresponding author; o.kolosov@lancaster.ac.uk, +44 (0)1524 593619

## 1. Experimental setup.

Figure S1 shows a photograph of the AFM setup at Daresbury Laboratory with the beam path from the FEL and the guiding mirrors plus focusing lenses. A beam splitter directed some of the light into a pyro-detector to measure the laser intensity for normalising of the spectra data later.

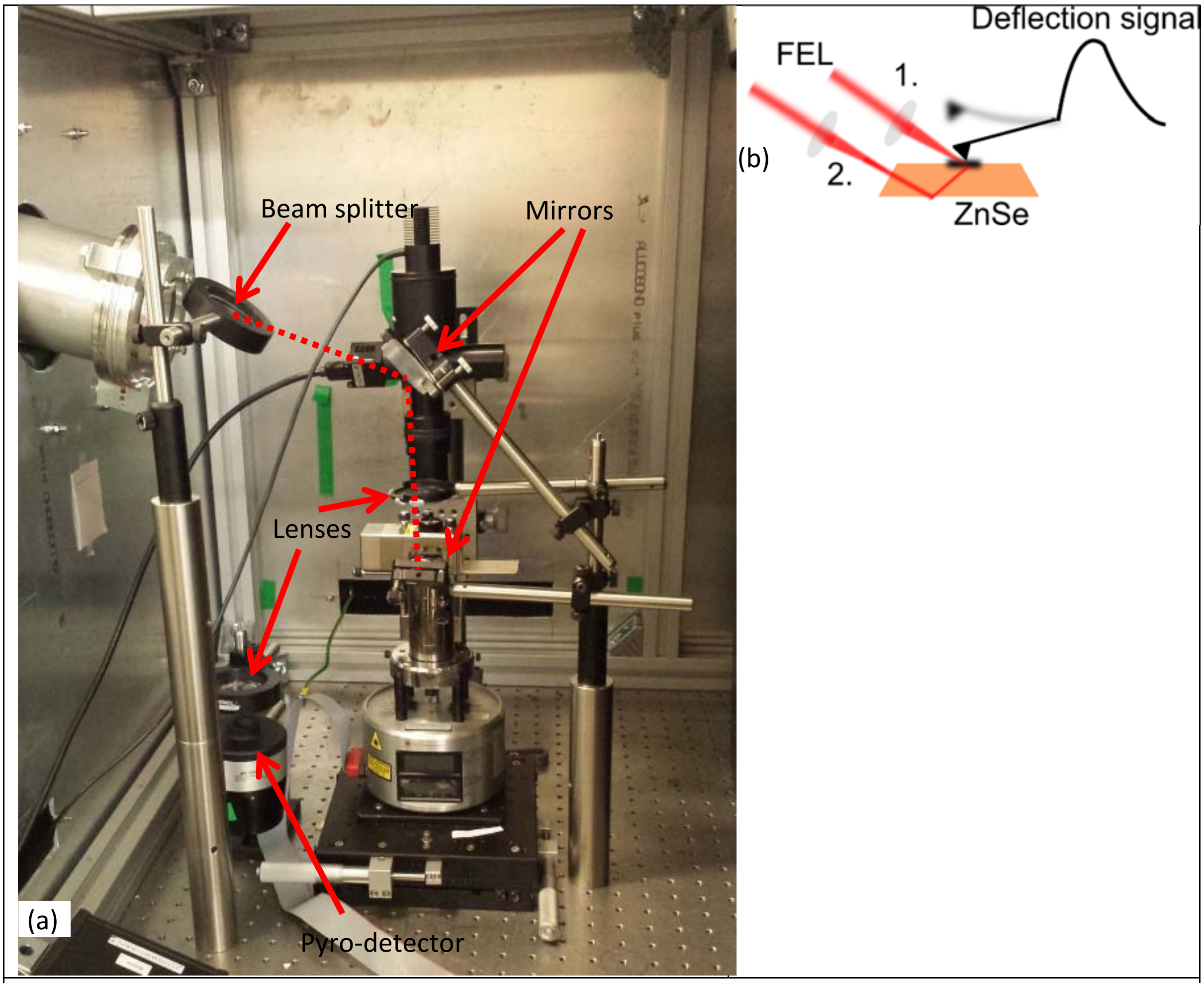

**Fig S1.** Photograph of experimental apparatus setup for AFM-IR at Daresbury Laboratories (a) and (b) schematic diagram showing the optics for i) The laser position for top down illumination and ii) Laser position for bottom up illumination. Dotted line in (a) shows the FEL light path.

An optical CCD camera connected to a laptop and a 10x objective was mounted above the AFM to allow for observation of cantilever position on the sample.

**1.1. *Aβ amyloid fibre sample purification and preparation:*** First the amyloid fibres were split from 1 mg vials to 0.5 mg aliquots using 0.01 % $NH_4OH$ at pH 10.6, with the peptide being brought into solution by vortexing and four lots of 30 seconds sonication. The 0.5 mg vials were dried by centrifugation under vacuum, then dissolved in trifluoroacetic acid (TFA) containing 4.5 % thioanisol at 1 mg/ml, vortexed and sonicated, followed by gentle drying under a nitrogen stream. Finally the deseeded protein was treated again at 1 mg/ml with 1,1,1,3,3,3-hexafluoro-2-propanol (HFIP) and briefly vortexed and sonicated. This peptide was then split into working aliquots and dried by centrifugation under a vacuum to give a final protein mass of 22.5 µg per sample. All working aliquots of peptide were stored at -20°C until future use.

Given the changes in Aβ protein structure during aggregation it was necessary to isolate the fibrils, which are predominantly β-sheeted, from any smaller aggregates, which are potentially representative of an α-helical structure. Given the monomers/oligomers are difficult to separate by centrifugation, an airfuge was employed for isolation of the fibrils [23]. Samples of 144 hour aggregated Aβ1:42/40 peptide were spun in a Beckman airfuge for 1 hour at 125,000 *g*. The supernatant was pipetted off and tested using ThT to ensure no fibrils remained in suspension. The pellets were re-suspended in 100 µl DI water before some further dilution using DI water to a concentration of approximately 1:30.

## 2. Experimental results.

Fig S2 shows some experimental oscilloscope results for AFM-IR and SThM-IR. Fig S2 (a) the background noise signal when no FEL beam was present shows low levels. Fig S2 (b) and (c) present the mechanical response of the SThM cantilever under laser excitation. This response is similar to the AFM-IR mechanical response. This is also compared with the SThM-IR thermal signal (Fig S2 (d)) for bottom up illumination displaying the same profile of the top down illumination results.

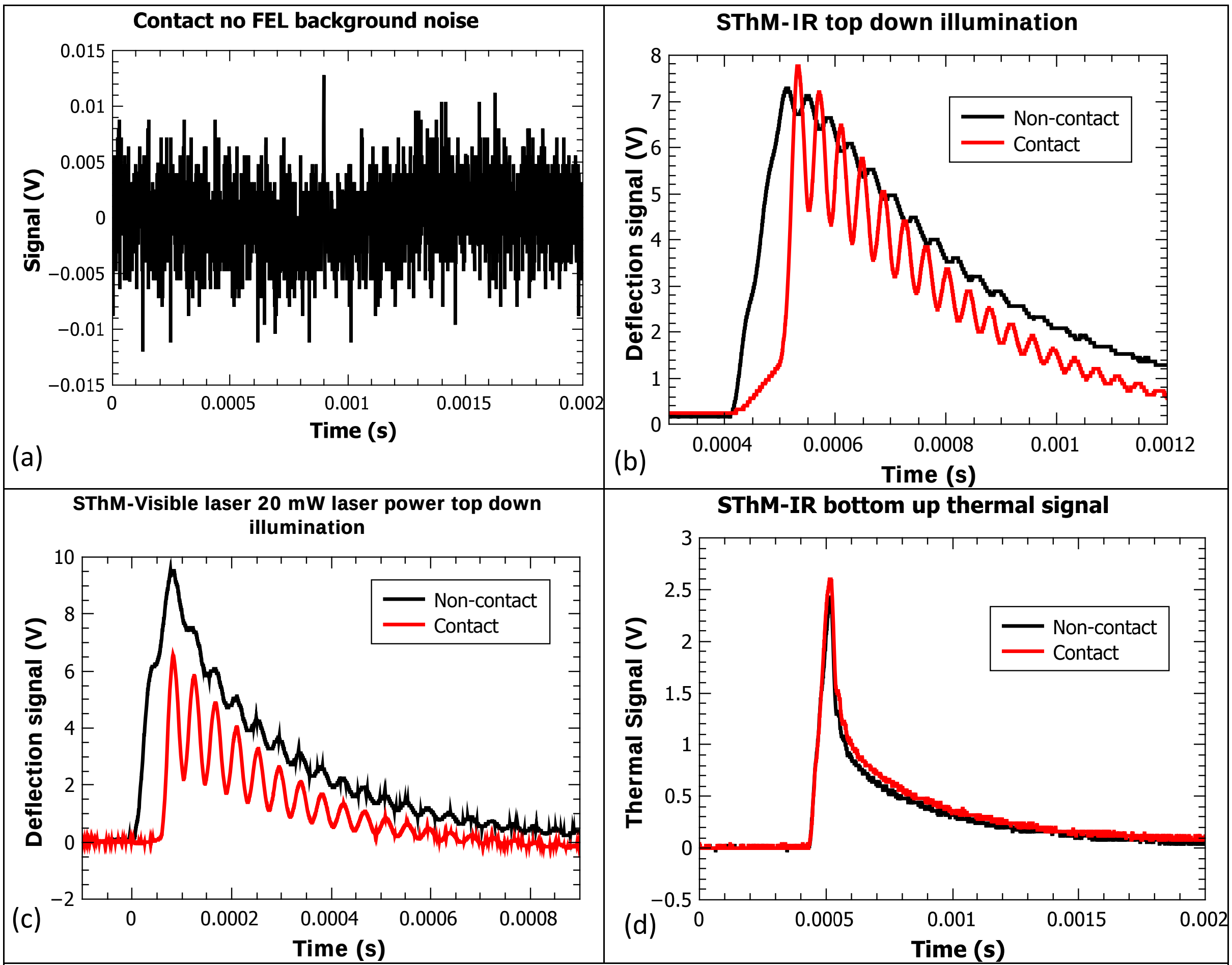

**Fig S2.** Oscilloscope read-out for (a) contact AFM-IR with no FEL beam showing background noise, (b) SThM-IR deflection response, (c) SThM-IR deflection response using the visible red laser and (d) the SThM-IR bottom up illumination thermal response.

The background noise level was very small and showed little variation compared with AFM-IR deflection signal strength. Both the IR and 637 nm laser had similar responses for top down and bottom up illumination. The 637 nm laser could also be modulated in power and pulse length with higher powers leading to higher temperature and deflection signals but with the sample ringing pattern. The increase in pulse time lead to higher signals and if longer than the temperature rise time the signal would reach a maximum before the laser pulse ended and then the signal decayed.

Fixed wavelength profile line scans on a sample of amyloid fibre on $CaF_2$ is shown in Fig. S3 giving both the raw data and the averaged profile for the corresponding figure 4 line scans.

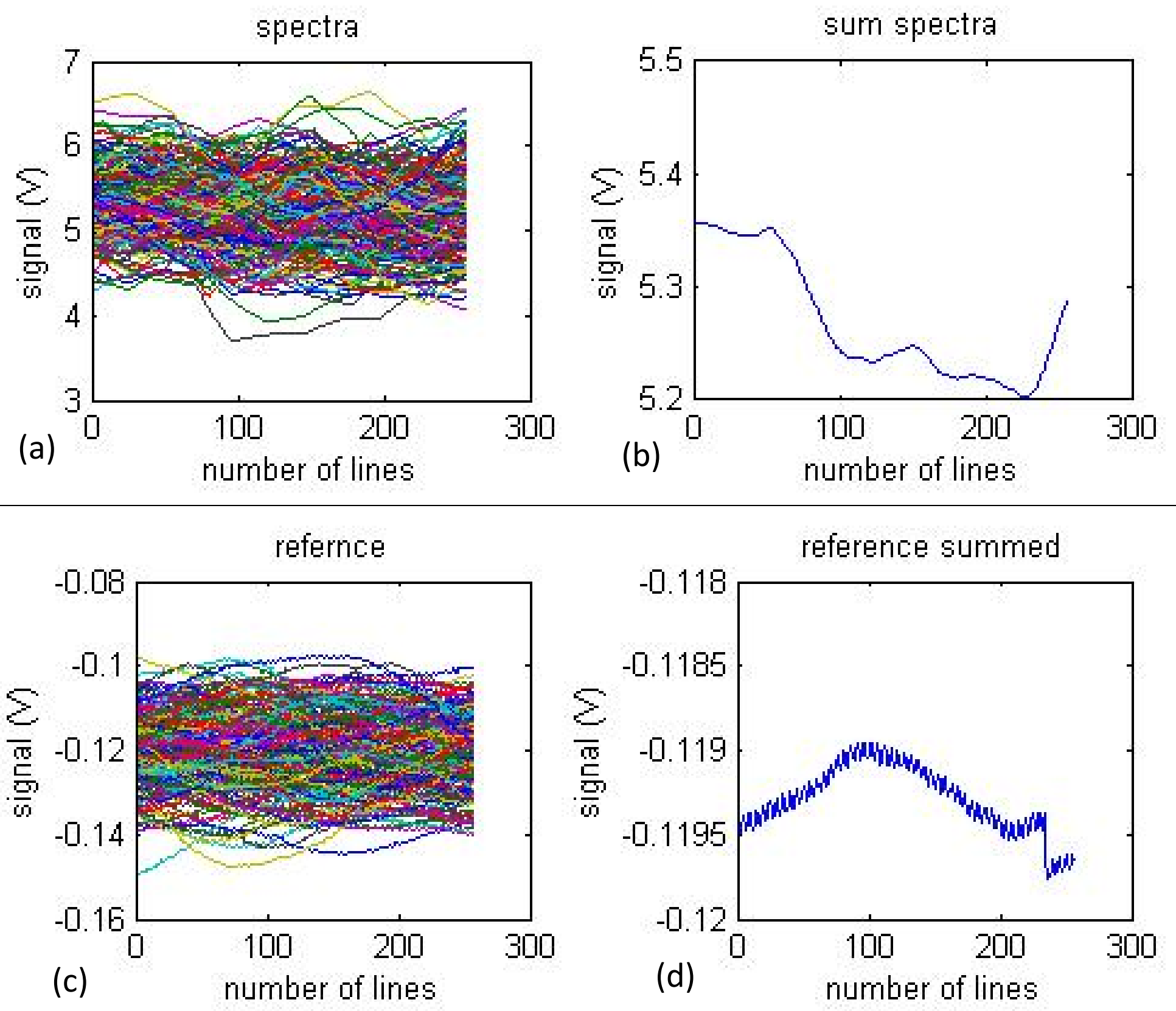

**Fig S3.** Line scan data for amyloid on Si substrate showing raw ((a) and (c)) and averaged ((b) and (d)) MATLAB outputs of the spectra ((a) and (b)) and the reference intensity ((c) and (d)).

The raw line scans were summed and averaged using the intensity signal from the detector. Figure S4 presents AFM-IR data taken with the Anasys NanoIR2 system on an amyloid fibres sample on gold substrate.

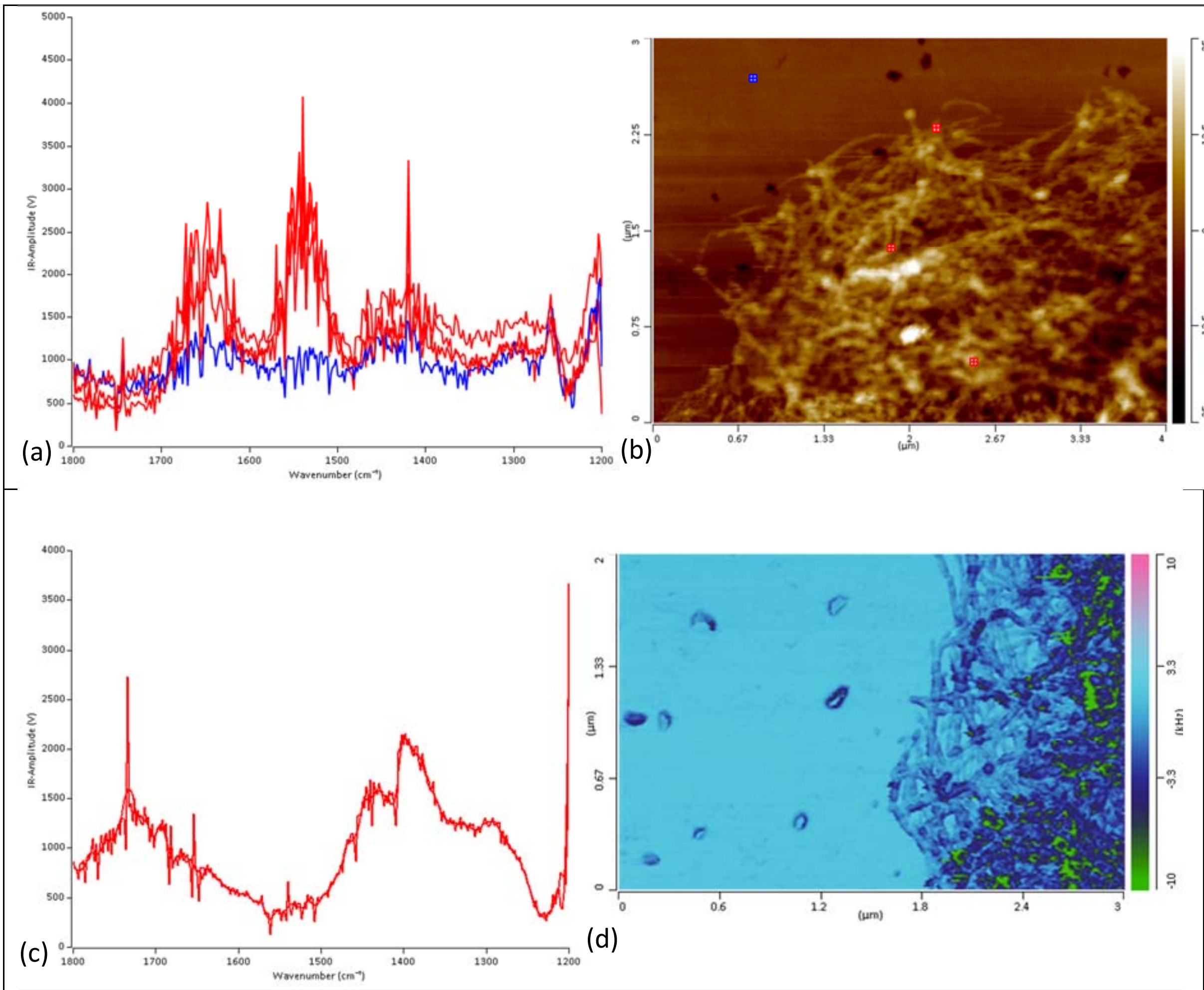

**Fig S4.** Anasys NanoIR2 unfiltered wavelength sweep (a) and AFM height image (b) on amyloid fibres on gold substrate along with (c) wavelength sweep on clean Au substrate and (d) frequency response image.

Spectra wavelength sweeps were done at different points on the amyloid fibres and gold substrate with the raw unsmoothed data shown in Fig. S4a, which clearly shows the amide I and II absorption peaks. The AFM-IR spectra response for a blank clean gold substrate also shows spectra absorption peaks. Figure S4 also displays the cantilever frequency response for a 2D AFM scan done at 1650 $cm^{-1}$ (Amide I).